\documentclass[reprint,aps,prl,groupedaddress,amsmath,amssymb]{revtex4-2}

\usepackage{hyperref}
\usepackage{xr, soul}
\usepackage{graphicx, dcolumn, xcolor, bm}

\newcommand{\dnn}{$l_\textrm{nn}$}
\newcommand{\mdnn}{$\langle l_\textrm{nn} \rangle$}
\newcommand{\vi}{ \mathbf{v}_{i} }
\newcommand{\mtau}{$\langle \tau \rangle$}
\newcommand{\mspd}{$\langle v_0 \rangle$}
\newcommand{\meps}{$\langle \epsilon \rangle$}
\newcommand{\mphi}{$\langle \Phi \rangle$}
\newcommand{\mlp}{$\langle l_p \rangle$}
\newcommand{\kexpr}{$\kappa = \langle l_p \rangle / \langle l_\textrm{nn} \rangle$}
\newcommand{\grpeak}{effective attraction}
\newcommand{\descriptors}{behavioural quantities}
\newcommand{\yy}[1]{{\color{black} #1}}
\newcommand{\smallfish}{young}
\newcommand{\bigfish}{old}
\usepackage[normalem]{ulem}

\begin{document}

\title{Dominating Lengthscales of Zebrafish Collective Behaviour}

\author{Yushi Yang}
	\affiliation{H.H. Wills Physics Laboratory, Tyndall Avenue, Bristol, BS8 1TL, UK}
	\affiliation{Bristol Centre for Functional Nanomaterials, University of Bristol, Bristol, BS8 1TL, UK}
\author{Francesco Turci}
	\affiliation{H.H. Wills Physics Laboratory, Tyndall Avenue, Bristol, BS8 1TL, UK}
\author{Erika Kague}
	\affiliation{Department of Physiology, Pharmacology, and Neuroscience, Medical Sciences, University of Bristol, Bristol, BS8 1TD, UK}
\author{Chrissy L. Hammond}
	\affiliation{Department of Physiology, Pharmacology, and Neuroscience, Medical Sciences, University of Bristol, Bristol, BS8 1TD, UK}
\author{John Russo}
	\affiliation{Department of Physics, Sapienza University, P.le Aldo Moro 5, 00185 Rome, Italy}

\author{C. Patrick Royall}
	\affiliation{Gulliver UMR CNRS 7083, ESPCI Paris, Université PSL, 75005 Paris, France.}
	\affiliation{H.H. Wills Physics Laboratory, Tyndall Avenue, Bristol, BS8 1TL, UK}
	\affiliation{School of Chemistry, Cantock Close, Bristol, BS8 1TS, UK}

\date{\today}

\begin{abstract}

Collective behaviour in living systems is observed across many scales, from bacteria to insects, to fish shoals. Zebrafish have emerged as a model system amenable to laboratory study. Here we report a three-dimensional study of the collective dynamics of fifty zebrafish. We observed the emergence of collective behaviour changing between  \yy{ordered} to randomised, upon \yy{adaptation} to new environmental conditions. We quantify the spatial and temporal correlation functions of the fish and identify two length scales, the persistence length and the nearest neighbour distance, that capture the essence of the behavioural changes. The ratio of the two length scales correlates robustly with the polarisation of collective motion that we explain with a reductionist model of self--propelled particles with alignment interactions.

\end{abstract}

\maketitle

\section*{Author Summary}

\yy{
Groups of animals can display complex collective motion, which emerges from physical and social interactions amongst the individuals. A quantitative analysis of emergent collective behaviour in animals is often challenging, as it requires describing the movement of many individual animals. With an innovative 3D tracking system, we comprehensively characterized the motion of large groups of zebrafish (Danio rerio), a freshwater fish commonly used as a vertebrate model organism. We find that the different collective behaviours are captured by two physical scales: the length of persistent motion in a given direction and the typical nearest neighbour distance. Their ratio allows us to interpret the experimental results, in the light of a statistical mechanics model for swarming with persistent motion and local neighbourhood alignment.
}

\section{Introduction}

In living systems aggregation occurs at different scales, ranging from bacteria (microns) to insects (centimetres) to fish shoals (tens of kilometres) and 
with emerging complex patterns \yy{\cite{liu2019prl, cavagna2010, makris2006}}. These manifestations of collective behaviour originate from the interactions among the individual agents and between the agents and the environment \cite{deneubourg1989}. Such interactions are often modelled by a combination of deterministic and stochastic contributions, capturing the individual's variability observed in nature and unknown or uncontrollable variables. The emergence of collective behaviour has been shown to be advantageous for communities \cite{pitcher1982, trenchard2016, hemelrijk2015}, \yy{and the identification of universal patterns across scales and species reveals the physics behind these phenomena} \cite{cavagna2017, mendez-valderrama2018}. Understanding the relationship between the collective behaviour and animal interactions has potential technological applications, for example to reverse engineer algorithms for the design of intelligent swarming systems \cite{grunbaum2005}. Successful examples include the global optimisation algorithm for the travelling salesman problem inspired by the behaviour of ants, and implementation of the Boids flocking model in schooling of robotic fish \cite{dorigo2006, jia2015}.

In a reductionist approach, collective behaviour can be modelled with interacting agents representing individuals in living systems. For example, groups of animals may be treated as if they were self-propelled particles with different interacting rules \cite{attanasi2014pcb, ni2016}. Examples of using simple agent--based models applied to complex behaviour include describing the curvature of the fish trajectories as a Ornstein--Uhlenbeck process \cite{gautrais2012}, modelling the ordered movement of bird flocks by an Ising spin model \cite{cavagna2010, bialek2012}, mapping of midge swarms onto particulate systems to explain the scale-free velocity correlations \cite{sinhuber2017, attanasi2014pcb, attanasi2014PRL} and swarming in active colloids \cite{bricard2013, yan2016}. One of the simplest approaches is the Vicsek model \cite{vicsek1995}, in which the agents only interact via velocity alignment. Despite its simplicity, a dynamical phase transition from \yy{ordered} flocking to randomised swarming can be identified, providing a basis to describe collective motion in biological systems \cite{couzin2002, vandervaart2019}.

The study of collective behaviour in living systems typically has focused on two-dimensional cases for reason of simplicity, making the quantitative characterisation of three-dimensional systems such as flocks of birds \yy{or} shoal of fish rare. To bridge this gap, zebrafish \textit{(Danio rerio)} present a wealth of possibilities \cite{giannini2020}: zebrafish manifest shoaling behaviour, i.e.~they form groups and aggregates, both in nature and in the laboratory; also, it is easy to constrain the fish in controlled environments for long--time observations. Typically, the response of fish to different perturbations, such as food and illumination, can be pursued \cite{giannini2020,puckett2018,jolles2017}. Furthermore, genetic modification has been very extensively developed for zebrafish, giving them altered cognitive or physical conditions, and yielding different collective behaviour \cite{kim2017, tang2020}.

However, tracking zebrafish in three dimensions (3D) has proven difficult \cite{pedersen2020}. To the best of our knowledge, previous studies on the 3D locomotion of zebrafish focussed either on the development of the methodology \cite{butail2011, wang2017}, or were limited to very small group sizes $(N \le 5)$ \cite{cachat2011, saberioon2016, pedersen2020}, while ideally one would like to study the 3D behaviour of a statistically significant number of individuals, representative of a typical community. In the field, zebrafish swim in 3D with group sizes ranging from tens to thousands \cite{shelton2020}.

Here we report on the collective behaviour of a large group ($N=50$) of wild-type zebrafish, captured by a custom 3D tracking system.
The observed fish shoals present different behaviours, showing different levels of local density and velocity synchronisation.
We identify two well-separated time scales (re-orientation time and state-changing time) and two important length scales (persistence length and nearest neighbour distance) for the zebrafish movement. The time scales indicate the fish group change their collective state gradually and continuously. The spatial scales change significantly as collective behaviour evolves over time, with strong correlations between spatial correlations and shoaling. Finally, we reveal a simple and universal relationship between the global velocity alignment of the shoals (the \textit{polarisation}) and the the ratio between the two length scales (the \textit{reduced persistence length}). We rationalise this finding through the simulation of simple agent-based models, in which an extra inertia term is added to the Vicsek model. Our findings illustrate complex behaviour in zebrafish shoaling, with couplings between spatial and orientational correlations that could only be revealed through a full three-dimensional analysis.

\section{Results}

\subsection{Experimental Observation}

\begin{figure*}
\centering
\includegraphics[width=0.7\linewidth]{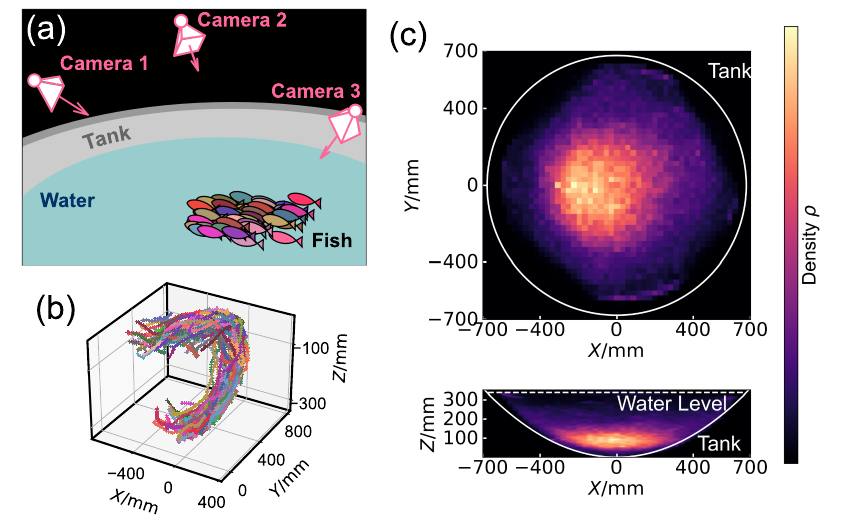}
\caption{\textbf{Experimental setup and overall spatial distributions.}
(a) Schematic illustration of the experimental setup. Zebrafish were transferred into a bowl-shaped tank and three cameras were mounted above the air-water interface to record the trajectories of the fish. 
(b) 3D trajectories obtained from the synchronised videos of different cameras. These trajectories belong to 50 {\smallfish} zebrafish (group Y1) in 10 seconds.
(c) The spatial distribution of 50 young fish (Y1) during a one-hour observation. Brighter colour indicates higher density. The top panel shows the result in XY plane, obtained from a max-projection of the full 3D distribution. The bottom panel shows a max-projection in the XZ plane. The outline of the tank and water-air interface, obtained from 3D measurement, are labelled.}
\label{figSpatial}
\end{figure*}

We tracked the movement of zebrafish from multiple angles using three synchronised cameras. We collected data for fish groups with different ages, with {\smallfish} fish (labelled as Y1--Y4) and {\bigfish} fish (labelled as O1--O4). Figure \ref{figSpatial}(a) schematically illustrates the overall setup of the experiment, where the cameras were mounted above the water to observe the fish in a white tank in the shape of a parabolic dish, enabling 3D tracking \cite{cavagna2010, kelley2013, stowers2017, ling2019}.
With this apparatus, we extract the 3D positions of the centre of each fish at different time points, with the frequency of 15 Hz. We then link these positions into 3D trajectories. Figure \ref{figSpatial}(b) presents typical 3D trajectories from 50 {\smallfish} zebrafish during a period of 10 seconds, where the fish group changed its moving direction at the wall of the tank.
The zebrafish always formed a single coherent group, without splitting into separate \yy{subgroups} during our observations. Movies are available in the published version~\cite{yang2022}. 
Figure \ref{figSpatial}(c) shows the cumulative spatial distribution of the zebrafish in the tank, during a one-hour observation. It is clear from this figure that the fish tend to swim near the central and bottom part of the tank. The propensity of zebrafish to swim near the wall was our motivation to use a bowl-shaped fish tank shown in(c), so that there are no corners for the fish to aggregate in, compared to a square-shaped container like conventional aquaria.

\subsection{Evolving Collective Behaviour}

\begin{figure*}
\centering
\includegraphics[width=0.8\linewidth]{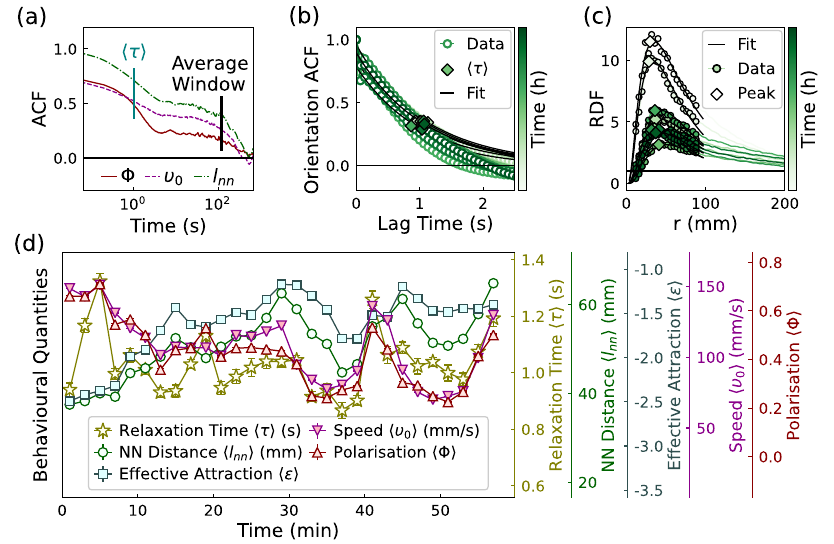}
\caption{\textbf{The {\descriptors} of 50 {\smallfish} zebrafish (group Y1).}
	(a) The auto--correlation function of the polarisation and average speed of the fish group.
	(b) The auto--correlation function of the orientations of fish.
	(c) Sequence of radial distribution functions with increasing time: at early times (top curves) the fish are clustered together so that the peak is large; at later times (bottom curves) the local density decreases and so does the peak height.
	(d) The time evolution of the averaged {\descriptors} for 50 {\smallfish} fish. Each point corresponds to the average value in 2 minutes.
	The error bars illustrate the standard error values.
}
\label{figTimeSeries}
\end{figure*}

The 3D tracking yields the positions of the fish, whose discrete time derivative gives the velocities. From these two quantities, we calculate three global descriptors to characterise the behaviour of the fish: the average speed, the polarisation, and the nearest neighbour distance. The average speed is defined as $v_0 = 1/N \sum{|\vi|}$ where $i$ runs over all the tracked individuals. The polarisation  $\Phi$ characterises the alignment of the velocities. It is defined as the modulus of the average orientation, written as \yy{\cite{attanasi2014pcb}},
$\Phi = 1/N \left| \sum{({\vi}/{| \vi |}}) \right|$ where $i$ runs over all the individuals. Large polarisation ($\Phi \sim 1$) signifies synchronised and ordered movement, while low polarisation indicates decorrelated, random movement. The nearest neighbour distance between the fish centres is defined according to the Euclidean metric, and we focus on is arithmetic mean {\dnn}.
\yy{These quantities were selected, because $v_0$ and $\Phi$ describe the dynamic of the fish, and {\dnn} captures structural information on the group of fish.}

We start from the analysis of temporal correlations of these three scalar quantities. Notably, all three exhibit two distinct time scales. Figure \ref{figTimeSeries}(a) shows the auto--correlation functions (ACF) of $v_0$, $\Phi$ and {\dnn}  averaged over the group of 50 {\smallfish} fish, calculated from a one hour observation. The ACFs present two decays and one intermediate plateau.
\yy{We identify the first decay ($\sim$1s) corresponding to the reorientation time of the zebrafish.}
This can be shown through the analysis of the autocorrelation of the orientations Fig.\ \ref{figTimeSeries}(b), which are characterised by an exponential decay with relaxation time {\mtau} close to $\sim$1s
\yy{. This value} is compatible with the previously reported turning rate timescale ($\sim$0.7s) \cite{mwaffo2015}.

The plateau and subsequent decay of the ACF of the scalar quantities $v_0$, $\Phi$ and {\dnn}, with the time scale of $\sim$120 seconds, represent complete decorrelation from the initial state, indicating that the shoal properties change significantly on this much longer timescale.
Therefore, we employ time-windows of 120 seconds to average the time evolution of of $v_0$, $\Phi$ and {\dnn}, to characterise the states of the fish groups with moving averages {\mspd}(t), {\mphi}(t) and {\mdnn}(t).

\begin{table*}[b]
\caption{\label{tabSymbol}
    A summary of the variables used to describe the fish behaviour.
}
\begin{ruledtabular}
\begin{tabular}{lccr}
    Symbol & Name & Unit & Comment \\
\colrule
    $v_0$ & Speed & mm/s & Average over different fish\\
    {\dnn} & Nearest Neighbour Distance & mm & Average over different fish\\
    $\Phi$ & Polarisation &  1 & Larger = ordered movement\\
    $\left\langle \cdot \right\rangle$ & Average operator  & &  Time average over 120 s \\
    {\mtau} & Relaxation time & s & The relaxation of fish orientation \\
    {\meps} & Effective Attraction & 1 & Smaller = more cohesive\\
    {\mlp} & Persistence Length & mm & Defined as {\mspd}{\mtau} \\
    $\kappa$ & Reduced Persistence Length & 1 & Defined as {\mlp} / {\mdnn} \\
\end{tabular}
\end{ruledtabular}
\end{table*}


To characterise the degree of spatial correlation of the fish, we 
\yy{calculate}
their radial distribution function (RDF), see Fig.\ \ref{figTimeSeries}(c), which quantifies the amount of pair (fish-fish) correlations and it is commonly employed in the characterisation \yy{of} disordered systems ranging from gas to liquids, from plasma to planetary nebulae\cite{hansen2013}. Details on the RDF can be found in the supplementary information (SI). All the RDFs exhibit one peak at a short separation, indicating the most likely short-distance separation between fish.
The peak height is a measure of the cohesion of the fish group. Inspired by liquid state theory \cite{hansen2013}, we take the negative logarithm of the peak height to characterise what we call as the ``{\grpeak}'' among the fish, noted as {\meps}. While {\dnn} quantifies a characteristic lengthscale in the macroscopic collective state, $\epsilon$ quantifies the fish propensity to remain neighbours. In Fig.~\ref{figTimeSeries}\yy{(d)} we see that \yy{\mdnn} and \yy{\meps} are strongly correlated, confirming that {\dnn} is also a measure of the cohesion of the collective states. We term {\mspd}, {\mphi}, {\mtau}, {\mdnn}, and {\meps} ``{\descriptors}'', and the brackets indicate the moving average. \yy{These variables are summarised in Table~\ref{tabSymbol}.}

Figure \ref{figTimeSeries}(d) illustrates the time-evolution of all the {\descriptors}, calculated from the movement of 50 {\smallfish} fish (group Y1) ten minutes after they were extracted from a husbandry aquarium and introduced into the observation tank. Over time, the {\descriptors} drift, indicating that a steady state cannot be defined over the timescale of 1 hour.
This result is generic, as the separated time scales and changing states were obtained from repeated experiments on the fish group (Y2--Y4), and also from different groups of {\bigfish}er zebrafish (O1--O4).

\subsection{Shoaling State Described by Two Length Scales}

\begin{figure*}
    \centering
    \includegraphics[width=0.9\linewidth]{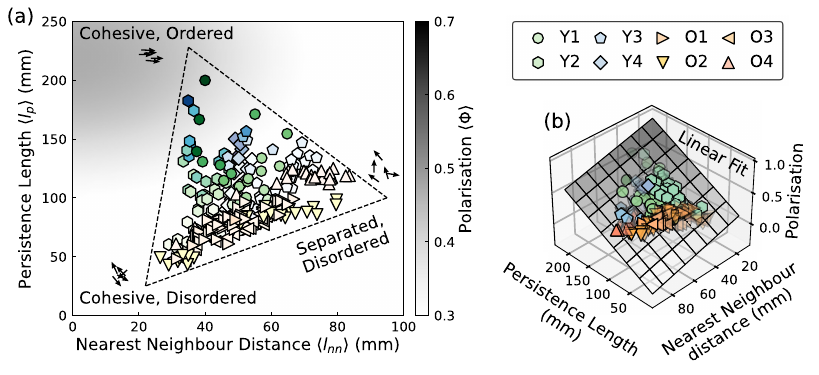}
    \caption{\textbf{The states of Zebrafish characterised by two length scales.}  (a) The states of the fish represented by the nearest neighbour distance and the persistence length. The brightness of the markers corresponds to the value of the polarisation. Each scatter--point corresponds to a time-average of 2 minutes. Different shapes indicate different fish groups from different experiments. (b) A multilinear regression model fitting the relationship between the polarisation and the two length scales, indicating the polarisation increase with the increase of persistence length, and the decrease of the nearest neighbour distance. The model is rendered as a 2D plane, whose darkness indicates the value of polarisation.}
    \label{figState2D}
\end{figure*}

To describe the space of possible collective macroscopic states we employ two dimensioned lengths, the nearest neighbour distance {\mdnn} defined above and a second scale characterising the typical distance that a single fish covers without reorientation, the persistence length $\langle l_p \rangle$ . \yy{This} is defined as the product of the speed and the orientational relaxation time $\langle l_p \rangle = \langle v_0 \rangle \langle \tau \rangle$. 

The resulting {\mlp} and {\mdnn} diagram is illustrated in Fig.~\ref{figState2D}(a). As we move across the diagram, the degree of alignment of the fish motion -- the polarisation -- also changes, indicating that changes in the local density (as measured by {\mdnn}) and in the pattern and velocity of motion (as measured by $\langle l_p \rangle$) are reflected in the
\yy{polarisation} of the shoals. For  high {\mlp} and low {\mdnn}, the movements of the fish are cohesive and ordered (Movie S1 in ~\cite{yang2022}). For the fish states with a low {\mdnn}  and low {\mlp}, the movements are cohesive but disordered (Movie S3 in ~\cite{yang2022}). For fish states with high {\mdnn} and low {\mlp}, the fish are spatially separated with disordered movements (Movie S2  in ~\cite{yang2022}). 
\yy{Separated and ordered} states are never observed. We also note that there is a systematic difference between young (Y) and old (O) fish groups, with the former characterised by
\yy{longer persistence lengths, shorter neighbour distances and larger polarisations,}
while the latter are clustered in a narrower range of persistence lengths with more disorder.

The simplest model to capture the relationship between polarisation and the two lengthscales is a multilinear regression. This yields $\langle \Phi \rangle = 0.039\ \langle l_p \rangle - 0.05\ \langle l_\textrm{nn} \rangle + 0.147$, with a goodness of fit value $R^2 = 0.73$.  This strong simplification suggests that most of the fish macroscopic states reside on a planar manifold in the $\Phi$--{\dnn}--$l_p$ space, illustrated in Fig.\ \ref{figState2D} (b). The value of {\mphi} increases with the increase of {\mlp}, and the decrease of {\mdnn}. Such relationship is reminiscent of results from the agent-based Vicsek model, where the polarisation of self--propelled particles is determined by the density ($\sim l_\textrm{nn}^{-1}$) and orientational noise ($\sim l_p^{-1}$) \cite{vicsek1995, ginelli2016}. In addition, the relationship between the polarisation and the local density suggests a metric based interaction rule, rather than the topological one \cite{ballerini2008}. In other words, the fish tend to align with nearby neighbours, rather than a fixed number of neighbours.
\yy{For instance, if the fish always align with their closest neighbours regardless of the distance, then the polarisation of the system will not be affected by {\dnn}}.
A similar relationship between polarisation and density was also found for jackdaw flocks while responding to predators \cite{ling2019nc}.

\yy{Interestingly, the ratio between the persistence length and the nearest neighbour distance exhibits a simple and robust correlation with the polarisation.}
Here we introduce the \emph{reduced} persistence length {\kexpr}. The value of $\kappa$ exhibits a consistent relationship with the polarisation for all the fish groups, as shown in Fig.\ \ref{figState1D} (a). All the experimental data points collapse onto a single curve, especially for the younger fish groups (Y1--Y4) which have a much wider dynamic range than the older groups. Notably, the young fish always transform from ordered states with high $\kappa$ value to disordered states with low $\kappa$ value, possibly because \yy{they adapt} to the observation tank.

To understand this relationship, we consider the fish motion as a sequence of persistent paths interrupted by reorientations. In a simplified picture, the new swimming direction at a reorientation event is determined by an effective local alignment interaction that depends on the neighbourhood, and notably on the nearest neighbour distance {\dnn}. The fish states with larger value of $\kappa$ correspond to situations where each individual fish interacts with more neighbours on average, between successive  reorientations. The increased neighbour number leads to a more ordered collective behaviour, so that the values of $\kappa$ and $\Phi$ are positively correlated as shown in Fig.\ \ref{figState1D}(a).

The time-averaged spatial correlation of the velocity fluctuation \yy{supports} our picture of the local alignment interaction between the fish. Such a correlation function is defined as,

\begin{equation}
C(r) = \frac{
\sum_{i=1}^{N}\sum_{j=i+1}^{N}
{ \left[
(\mathbf{v}_i - \bar{\mathbf{v}}) \cdot (\mathbf{v}_j - \bar{\mathbf{v}})\ 
\delta( r - r_{ij})
\right] }
} {
\sum_{i=1}^{N}\sum_{j=i+1}^{N}
\delta( r - r_{ij})
},
\end{equation}
where $\vi$ is the velocity of fish $i$, $\bar{\mathbf{v}}$ is the average velocity in one frame, $r_{ij}$ is the distance between two
\yy{fish}, and $\delta$ is the Dirac delta function. This function is widely used to characterise the average alignment of velocity fluctuations of moving animals, at different distances \cite{cavagna2010, silverberg2013, vandervaart2020}. Figures \ref{figState1D}(b) and (c) show the correlation functions for different fish groups with low and high $\kappa$ values, respectively. The distances are rescaled by the different {\mdnn} values of each fish group. For both conditions, the correlation curve collapses beyond one {\mdnn}, and peaks around the value of {\mdnn}, supporting our assumption that {\mdnn} is the length scale for the fish--fish interactions.

\subsection{Vicsek Model Rationalisation of the Experiments}

\begin{figure*}
\centering
\includegraphics[width=\linewidth]{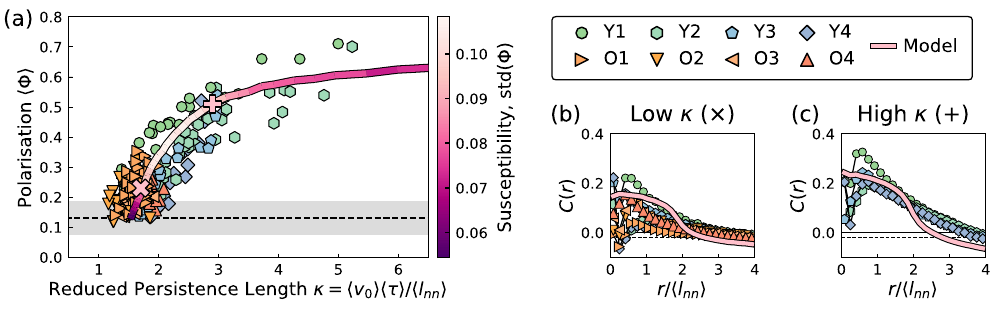}
\caption{\textbf{Single--parameter description of the Zebrafish system.}
(a) The average polarisation {\mphi} as a function of the reduced persistence length $\kappa$, where most data points collapse, and agree with the simulation result of the inertial Vicsek model. The dashed line and grayscale zone represent the expected average value and standard deviation of {\mphi} for the uniform random sampling of vectors on the unit sphere.
(b) The velocity correlation function of the fish and the model in the low $\kappa$ states, highlighted in (a) with a $\times$ symbol.
(c) The velocity correlation function of the fish and the model in the high $\kappa$ states, highlighted in (a) with a $+$ symbol.}
\label{figState1D}
\end{figure*}

The relationship between $\kappa$ and $\Phi$, presented in Fig.~\ref{figState1D}(a), can be easily compared with simulations. Here we explore this through simulations proposing a new modification of the original Vicsek model \cite{vicsek1995}. The Vicsek model treats the fish as point-like agents with an associated velocity vector of constant speed $v_{0}$. During the movement, the fish adjust their orientations to align with the neighbours' average moving direction. In order to take into account of memory effects in a simple fashion, we add an inertia term into the Vicsek model, so that each agent partially retain their velocities after the update, with the following rule
\begin{equation}
\footnotesize
\mathbf{v}_{i}(t+1) = v_0 \Theta \left[(1 - \alpha)\; \underbrace{
v_{0} \mathcal{R}_{\eta}
\left[\Theta\left(
    \sum_{j \in S_{i}} \mathbf{v}_{j}(t)
\right)\right]
}_\text{Vicsek Model}
+ \alpha \mathbf{v}_i(t)
\right],
\label{eqIVM}
\end{equation}
where $\vec{v}_i$ is the velocity or the $i$th fish, and the updated velocity of fish $i$ is a linear mixture of its previous velocity and a Vicsek term. The parameter $\alpha$ characterises the proportion of the non-updated velocity, \emph{i.e.} the inertia. This model is reduced to the Vicsek model by setting $\alpha$ to 0. If $\alpha = 1$, these agents will
\yy{perform straight motion with constant speed}
without any interaction. For the Vicsek term, $S_{i}$ is the set of the neighbours of fish $i$, and the $\Theta$ is a normalising function. The operator $\mathcal{R}_\eta[\mathbf{r}]$ randomly rotates the vector $\mathbf{r}$ into a new direction, which is drawn uniformly from a cap on the unit sphere. The cap is centred around $\mathbf{r}$ with an area of $4\pi\eta$. The value of $\eta$ determines the degree of stochasticity of the system. Our model is thus an inertial Vicsek model in three dimensions with scalar noise.

We set the units of the interaction range $\xi$ and time $dt$ and fix the number density to $\rho = 1\xi^{-3}$ and speed $v=0.1 \xi/dt$. We then proceed with varying the two parameters $\alpha$ and $\eta$ to match the data. In particular we measure the average persistence length $\langle k\rangle$ and polarisation {\mphi} and find that for $\alpha=0.63$ we can fit the data only through the variation of the noise strength $\eta$ (more details of the simulation are available in the SI). For $\eta \sim 1$, the movement of each agent is completely \yy{disordered}, reproducing the low $\kappa$ (or $\Phi$) states of the fish. For the case of $\eta \sim 0.65$ the movements of the agents are ordered ($\Phi \sim 0.64$) and mimic the states of fish with high $\kappa$. This is consistent with the fact that in the ordinary Vicsek model the persistence length scales as $\ell\sim v_0/\eta^2$ \yy{(Fig.~S7) \cite{ginelli2016}}.
The good fit of the simulation result suggests the fish--fish alignment interaction dominates their behaviour, and the fish can change their states by changing the rotational noise ($\eta$).

We emphasise that the inertial Vicsek model is a crude approximation, as the only interaction of the model is velocity alignment. Without the attractive/repulsive interactions and other details, the inertial Vicsek model does not reproduce the velocity correlation function of the fish, as illustrated in Fig.\ \ref{figState1D}(b) and(c), suggesting that more sophisticated models with effective pairwise and higher order interactions may be developed in the future. Nevertheless, the model qualitatively reproduces the fact that the velocity correlation is stronger in the high $\kappa$ states.

\section{Discussion and Conclusion}

Our results confirm some previous observations and open novel research directions. The {\smallfish} fish appear to adapt to a new environment with the reduction of the effective attraction and speed (Fig.\ \ref{figTimeSeries}). Such behaviour is consistent with previous observations of dense groups of fish dispersing over 2-3 hours \cite{delaney2002},
\yy{and it might be related to the fact that the fish perceived less danger as they adapted to the new environment \cite{miller2012, perez-escudero2017}.}
At the same time, it was reported by Miller and Gerlai that the habituation time has no influence on the Zebrafish group density \cite{miller2007}. We speculate that this difference emerges from the way the statistics were performed. Typically, Miller and Gerlai's results were averaged over 8 different small fish groups (N=16), and it is possible that the noise among different groups obfuscates the time dependence features here highlighted. Despite the different claims, our result matched Miller and Gerlai's result quantitatively (Fig.~S8).

\yy{
The strong correlation between the speed and polarisation (Fig.~\ref{figTimeSeries}(d)) is consistent with previous studies on zebrafish \cite{miller2013, heras2019}. Such correlation had also been observed for different fish species \cite{gautrais2012, tunstrom2013, tang2020}.
The polarisation of the {\smallfish} fish was found to decrease (Fig.~\ref{figTimeSeries}(d)), during the adaptation process.}
This ``schooling to shoaling'' phenomenon \yy{had also} been observed previously in a quasi 2 dimensional environment \cite{miller2012}.
Our results\yy{, being quantitatively consistent with previous reports (Fig.~S9),}
suggest that this behaviour is present also in a fully three-dimensional context and that the change from schooling  (\yy{ordered} motion) to shoaling (\yy{disordered} motion) is related to an increasingly disordered or uncorrelated behaviour, corresponding to the increase in the noise term $\eta$ in the Vicsek model.

It is been speculated that all the biological systems were poised near the critical state, \yy{enjoying} the maximum response to the environmental stimuli \cite{mora2011}. Here the inertial Vicsek model \yy{offers} a supporting evidence \yy{to this claim}. The fluctuation of the polarisation, the \emph{susceptibility}, took a maximum value at moderate reduced persistence value $\kappa \sim 2$, as illustrated Fig.\ \ref{figState1D} (a). \yy{Also,} the fish states were clustered around such region, where the fish can switch between the disordered behaviour and ordered behaviour swiftly. Such disordered but critical behaviour was also observed for the midges in the urban parks of Rome\cite{attanasi2014PRL}.

In conclusion, our work presents a quantitative study of the spatial and temporal correlations manifested by a large group of zebrafish. In our fully 3D characterisation, we have shown that there is a timescale separation between rapid reorientations at short times and the formation of a dynamical state with characteristic spatial correlations at longer times. Such spatial correlations evolve continuously and no steady state is observed in the time window of one hour. Our analysis shows that the continuously changing collective macroscopic states of the fish can be described quantitatively by the persistence length and nearest neighbour distance. The ratio of these length scales presents a characteristic correlation with the polarisation of the fish group. This simple relation is supported by an elementary agent based model in the class of the Vicsek models for collective behaviour.

Our analysis also opens multiple questions: the true nature of the interactions and how these are linked to the sensory and vision capabilities of the fish is open to debate; also, the reason for the change of the fish states remained unexplored, with the possibility of the fish learning over time about the experimental conditions. Our work shows that zebrafish provides a viable model system for the study of animal collective behaviour where such questions can be investigated in a quantitative manner.

A further intriguing possibility is to link the methodology that we develop here, with genetic modifications to zebrafish, for example with behavioural phenomena such as autism \cite{kim2017} or physical alterations such as the stiffened bone and cartilage \cite{lawrence2018}.

\section{Methods}

\subsection{Data and Code Availability}

The code for tracking the fish, including the 2D feature detection, 3D locating, trajectory linking, and correlation calculation, is available free and open--source. \yy{It is available on GitHub: (\href{https://github.com/yangyushi/FishPy}{https://github.com/yangyushi/FishPy})}. The simulation code is also available on GitHub \yy{(\href{https://github.com/yangyushi/FishSim}{https://github.com/yangyushi/FishSim})}. The dataset for generating Fig.\ \ref{figTimeSeries}, \ref{figState2D}, and \ref{figState1D} are available as supplementary information (Dataset S1), as well as some pedagogical code snippets (Code S1).

\subsection{Zebrafish Husbandry}

Wildtype zebrafish were kept in aquarium tanks with a fish density of about \yy{5} fish / L. The fish were fed with commercial flake fish food (Tetra Min). The temperature of the water was maintained at 25\textdegree C and the pH $\approx$ 7. They were fed three times a day and experience natural day to night circles, with a natural environment where the bottom of the tank is covered with soil, water plant, and decorations as standard conditions \cite{westerfield2000}. Our {\smallfish} group (Y) were adults between 4-6 months post-fertilisation, while our {\bigfish} group (O) were aged between 1-1.5 year. The standard body lengths of these fish were are available in the SI. All the fish were bred at the fish facility of the University of Bristol. The experiments were approved by the local ethics committee (University of Bristol Animal Welfare and Ethical Review Body, AWERB) and given a UIN (university ethical approval identifier).

\subsection{Apparatus}

The movement of the zebrafish were filmed in a separate bowl-shaped tank, which is immersed in a larger water tank of 1.4 m diameter. The radius $r$ increasing with the height $z$ following $z = 0.2 r^2$. The 3D geometry of the tank is measured experimentally by drawing markers on the surface of the tank, and 3D re-construct the positions of the markers. Outside the tank but inside the outer tank, heaters and filters were used to maintain the temperature and quality of the water. The videos of zebrafish were recorded with three synchronised cameras (Basler acA2040 um), pointing towards the tank. \yy{Detailed information is available in the SI}.

\subsection{Measurement and Analysis}

Fifty zebrafish were randomly collected from their living tank, moved to a temporary container, then transferred to the film tank. The filming started about 10 minutes after fish were transferred. The individual fish in each 2D images were located by our custom script and we calculated the 3D positions of each fish following conventional computer vision method \cite{hartley2003, cavagna2008AB}. 

The 3D positions of the fish were linked into trajectories \cite{ouellette2006, xu2008}. Such linking process yielded the positions and velocities of different fish in different frames. We segmented the experimental data into different sections of 120 seconds, and treat each section as a steady state, where the time averaged {\descriptors} were calculated. More details on the tracking and analysis are available in the SI.

\section*{Supplementary Information}

\title{Dominating Lengthscales of Zebrafish Collective Behaviour}

\author{Yushi Yang}
	\affiliation{H.H. Wills Physics Laboratory, Tyndall Avenue, Bristol, BS8 1TL, UK}
	\affiliation{Bristol Centre for Functional Nanomaterials, University of Bristol, Bristol, BS8 1TL, UK}
	
\author{Francesco Turci}
	\affiliation{H.H. Wills Physics Laboratory, Tyndall Avenue, Bristol, BS8 1TL, UK}
	
\author{Erika Kague}
	\affiliation{Department of Physiology, Pharmacology, and Neuroscience, Medical Sciences, University of Bristol, Bristol, BS8 1TD, UK}
	
\author{Chrissy L. Hammond}
	\affiliation{Department of Physiology, Pharmacology, and Neuroscience, Medical Sciences, University of Bristol, Bristol, BS8 1TD, UK}
	
\author{John Russo}
	\affiliation{Department of Physics, Sapienza University, P.le Aldo Moro 5, 00185 Rome, Italy}
	
\author{C. Patrick Royall}
	\affiliation{Gulliver UMR CNRS 7083, ESPCI Paris, Université PSL, 75005 Paris, France.}
\maketitle
\date{\today}

\section{Details of the Tracking System}

\subsection{The Apparatus}

\begin{figure*}
\centering
\includegraphics[width=120mm]{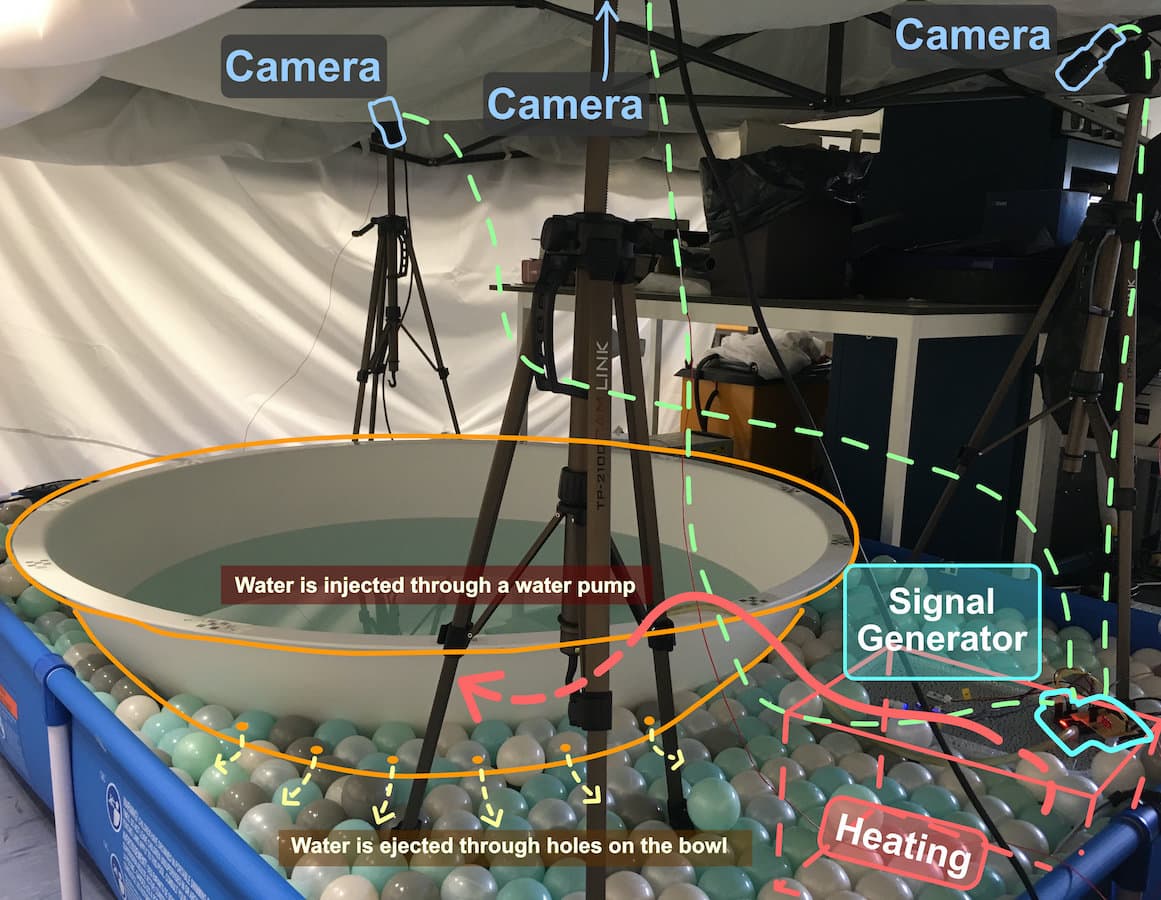}
\caption{A photo showing the 3D tracking apparatus featuring the cameras and the bowl--shaped fish tank. The cameras were triggered with signals generated by a Arduino microcontroller.}
\label{figLab}
\end{figure*}

Figure \ref{figLab} shows the photo of the apparatus where the large bowl shaped tank is placed inside a paddling pool, overlooked by three cameras. The cameras were triggered by an Arduino microcontroller, to generate synchronised videos using software Pylon Viewer. Our approach is different from a frequently used arrangement in which two orthogonal cameras were placed above and along side a transparent fish tank \cite{cachat2011, mwaffo2017, rosa2020}. This choice is motivated by the relatively large group size (50 fish). To determine 3D trajectories, we need to explicitly consider the optical details such as the distortion of the lens and the refraction of the water \cite{parrish1997}. Our setup mitigated the effects of refraction, since the light path from the fish to the camera is only refracted once by the water-air interface.

The entire system were heated by two commercial heaters. The warm water was circulated into the inner bowl with a water pump. The bowl was able to exchange water with outside via small holes drilled inside. The measured temperature in the swimming pool ranges from 23 \textdegree C to 26 \textdegree C. The water circulation is turned off during observations of fish swimming.

\begin{figure*}
\centering
\includegraphics[width=100mm]{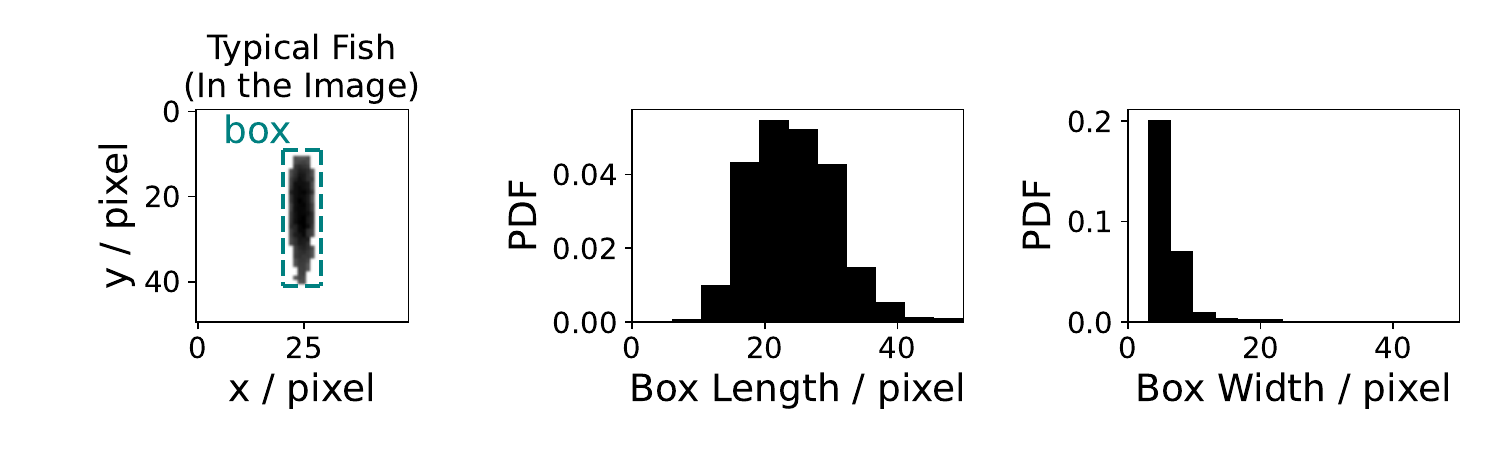}
\caption{The shape of fish captured by the cameras. Left: a typical fish shape in the image, and its bounding box. Centre: the distribution of the length of the bonding box. Right: the distribution of the width of the bonding box. The tail of the distribution is because of the bending of the fish body.}
\label{figFish}
\end{figure*}

The image size produced by our camera (acA2040 um, Basler) is 2056 pixels $\times$ 1540 pixels. We mounted a 6 mm fixed focal length lens (C Series, Edmund Optics) on the camera, which yields a wide view, allowing us to place the camera closer to the fish group. The typical distance between the camera and the fish group is around 2 meters. With such setup, the fish (body length $\sim$ 30 mm) appear as black rods in the videos. Figure~\ref{figFish} shows one typical shape of the fish, and the distribution of the fish sizes.

\subsection{Camera Calibration and Water Refraction}

\begin{figure*}
\centering
\includegraphics[width=120mm]{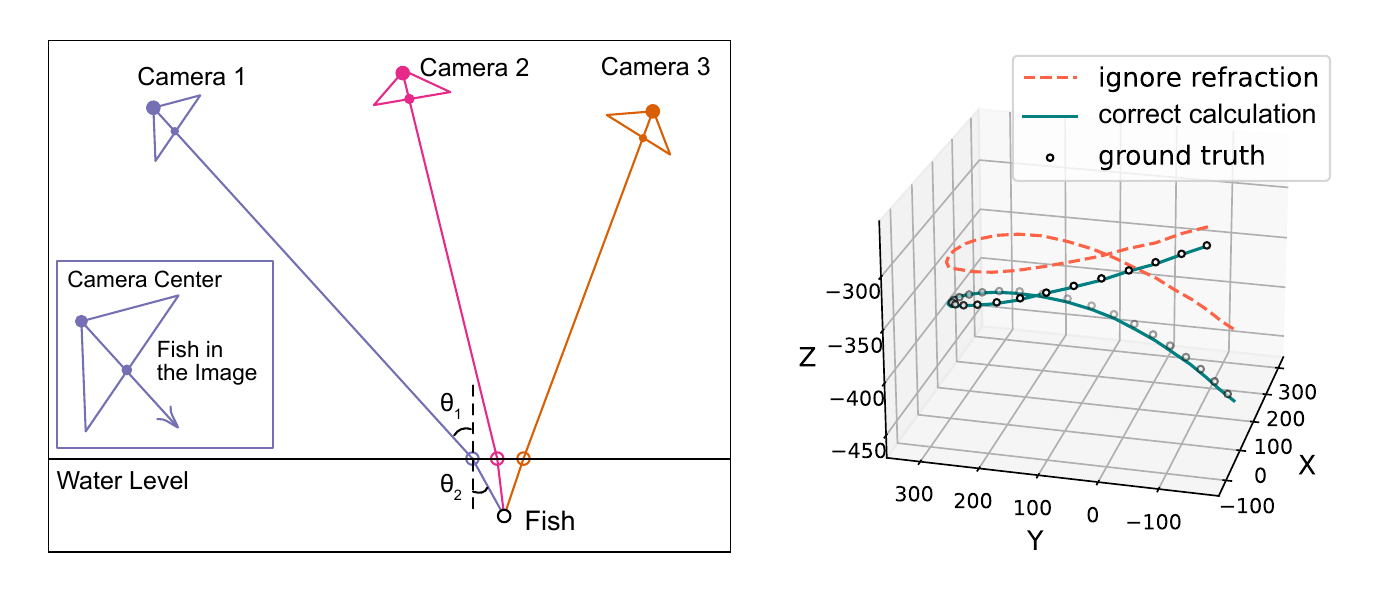}
 \caption{\label{figRefract}
The air-water interface.
Left: the schematic for the calculation of the 3D fish location, with the index change being considered.
Right: the effect of the refraction revealed by tracking simulated fish under water.
The solid line is the result where the refraction is considered, and the dashed line is the calculation result that ignored the refraction effect. The scatters were ground truth data that were used to generate the simulated image.}
\end{figure*}

The intrinsic parameters, including the camera matrix and the distortion coefficients, were obtained by common camera calibration procedure with the help of a chessboard. The intrinsic parameters and extrinsic parameters of the cameras were determined with a chessboard image floating on the water. By doing so, the origin of the coordinate system in the system global frame of reference was located at $(x, y, 0)$ and the air-water interface is fixed at $z=0$. This information is used to correct the water-refraction, when we are reconstruct the 3D positions of the fish. From the calibration, we also obtained the intrinsic camera matrices and the distortion coefficients.Figure~\ref{figRefract} (left) shows our method to take account of the air-water interface. With the extrinsic parameters of the cameras, we can calculate the centres of the cameras (in the global frame of reference). Detecting individual fish in the image, we can recast the light path responsible for its formation, visualised as the arrow in the insert of Fig.~\ref{figRefract} (left). Following the light path, we calculate its intersection with the air-water interface (the plane $z = 0$) and the direction of the light after the refraction following the Snell's law ($n_1 \sin\theta_1 = n_2 \sin\theta_2$). With three cameras, we collect the light from different angles. The intersection of these directions is the location of the fish.

Since our setup is relatively deep (300--400 mm in depth), the effect of the refraction can not be ignored. In Fig.~\ref{figRefract} (right) we showed the consequence of ignoring the refraction, which is a significant deviation from the true trajectory.

\subsection{Tracking Software}

The images were loaded using relevant functions from library ``opencv'', and the fish were separated from the static background by our custom imaging processing script. The location of fish in different videos were found by our custom 2D tracking code. These 2D locations were then used to calculate the 3D locations of the fish, utilising information about the calibrated camera. Typically, we explicitly considered the refraction of light by the water-air interface, knowing its exact level from the camera calibration procedure. We then link the positions into trajectories using a four-frames predictive procedure \cite{ouellette2006}. The obtained trajectories were furthered extended into longer ones following Xu's method \cite{xu2008}. All of the aforementioned procedures, including the image processing, 2D feature selection, 3D location and linking, are publicly available in Yushi's GitHub page \cite{yushi2021}.

\begin{figure}
\centering
\includegraphics[width=\linewidth]{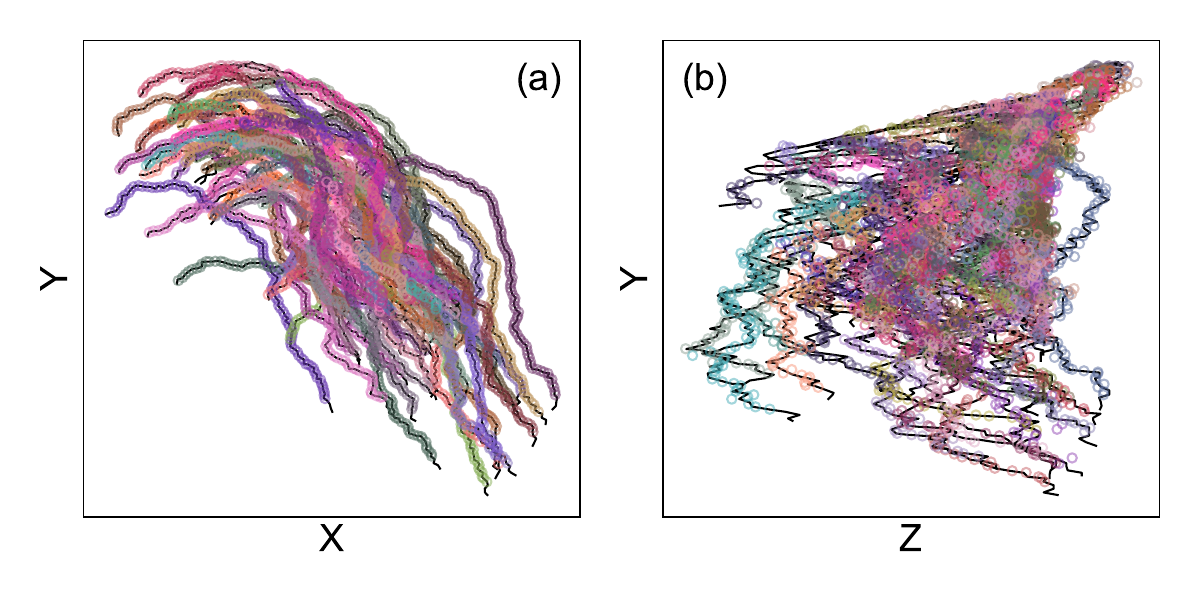}
\caption{The tracks of the rendered simulation with the corresponding ground truth. (a) The trajectories projected on XY plane. (b) The trajectories projected on YZ plane. The circles are the tracking results, and the black lines are the ground truth.}
\label{figBenchmarkTrajs}
\end{figure}

\subsection{The Accuracy of the Tracking Result}

The accuracy of our 3D tracking procedure is evaluated by tracking a rendered fish animation, where the fish were simulated as Vicsek models constrained inside a fish tank. The sizes of the fish and the tank were selected to be close to the experiments. We used the ``cycles'' engine in software ``blender'' (version 2.91.0 on Ubuntu 20.04) to render the animation, because the ray-tracing renderer can accurately model the refraction of water.

\begin{figure*}
\centering
\includegraphics[width=120mm]{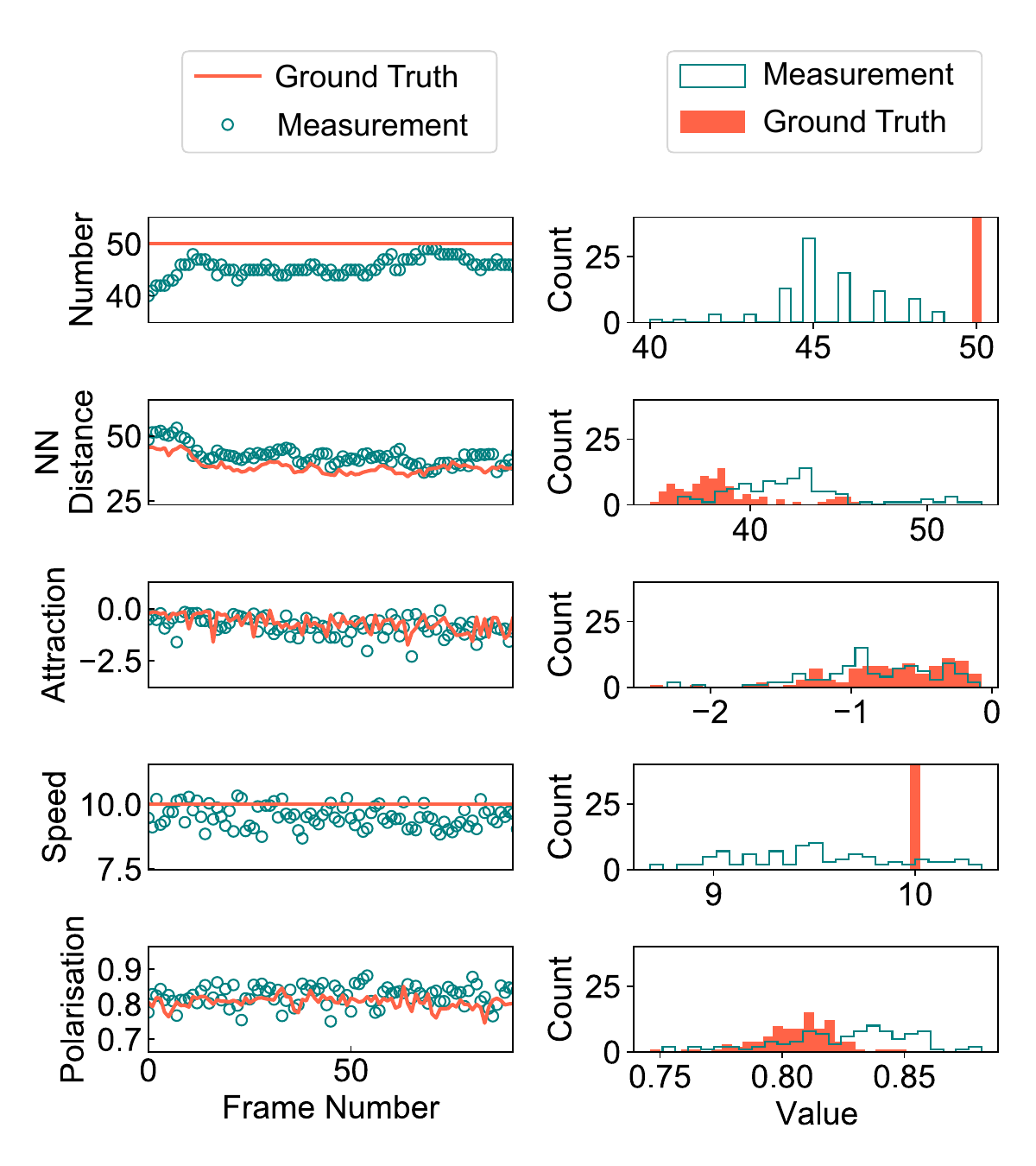}
\caption{Comparing the measured behavioural quantities with their corresponding ground truth values. From top to bottom are, the number of tracked fish; the average nearest neighbour distance; the average speed; the polarisation; and the effective attraction. These quantities were calculated frame-by-frame without time averaging. Left: the time evolution. Right: the histogram of different quantities.}
\label{figBenchmark}
\end{figure*}

The measured trajectories are very close to the ground truth (the simulation result), as illustrated in Fig.\ \ref{figBenchmarkTrajs}. The main error from the tracking is the missing of fish from frame to frame. That is, one fish might be tracked in frame $i$, but it is not tracked in frame $i+1$. The situation is illustrated in Fig.\ \ref{figBenchmark} (a), where only about 90\% ($\approx$45) of the fish were tracked on each frame. As a direct consequence, the measured nearest neighbour distance (\mdnn) is consistently larger than the ground truth (Fig.\ \ref{figBenchmark} (b)). The {\grpeak}, as an alternative quantity which is related to the local density, was found to be more robust, because is it relatively insensitive to any missing coordinates.

\section{The Age of the Fish}

\begin{figure}
\centering
\includegraphics[width=\linewidth]{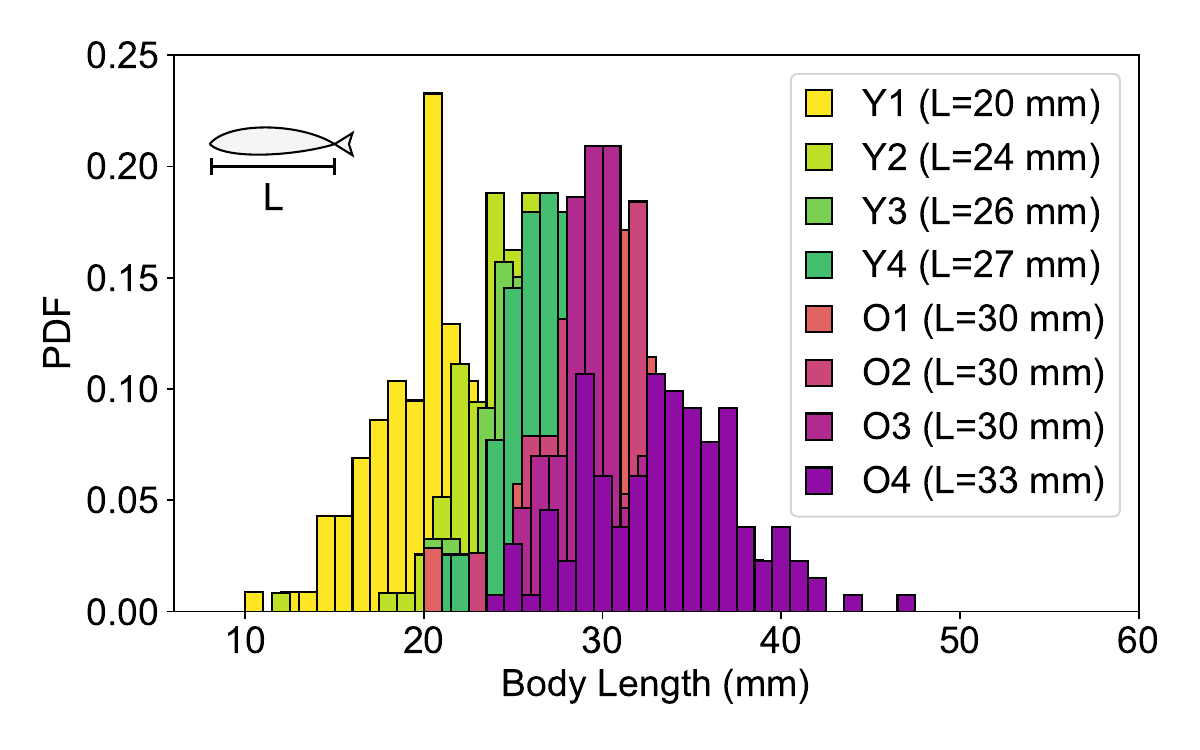}
\caption{The distribution of standard body length of different fish groups. The body lengths were measured from the videos manually, with a custom software. The average body length were included in the legends.}
\label{figBodyLength}
\end{figure}

We determine the age of the fish by measuring their standard body lengths. The results are illustrated in Fig.\ \ref{figBodyLength}. For group Y1--Y4, we performed the experiment for the same group four times, between August to November in the year of 2019. The fish grew in size during these time as their average body lengths increased from 19.6 mm to 26.5 mm. For group O1--O3, the fish were over 1 years old with the average body length of 30 mm. We performed the experiment for the same group 3 times, in February, 2021. The fish in group O4 were from another different group who are over 1 years old with the average body length of 33.5 mm. The experiments for the O4 group were performed in August, 2019.

\section{Details of the Analysis}

\subsection{Obtaining the positions and velocities}

We obtained the positions and velocities of different fish from the linked trajectories. The 3D positions that did not belong to any trajectory were discarded. In our linking process, we allow fish to disappear in some frames, creating voids in the trajectories, which were filled by linear extrapolation.

We segment the fish movement into different sections, of duration 120 seconds, corresponding to a state of the fish group. For each section, we calculated the value of $\Phi$, $d_1$ and $v_0$ frame-by-frame, and take their average to characterise the state for this section. For the value of {\meps} we calculated them from the radial distribution function. For the value of {\mtau}, we collected the trajectories inside the section, and clipped the trajectory if it extended outside the section of interest. Then we calculate the auto-correlation function of the orientations $C_o(t)$ as
$ C_o(t) = \left\langle \mathbf{v}_0(i) \cdot \mathbf{v}_0(i + t) \right\rangle, $
\noindent where $\mathbf{v}_0 = \mathbf{v} / |\mathbf{v}|$ is the velocity orientation of the fish, and the average is taken over all possible frame numbers ($i$). The initial decay of $C_o(t)$ is fitted with an exponential decay function $\exp(-t / b)$, by weighting the fitted data by the magnitude of $t$, and the fitting parameter $b$ is taken as {\mtau}.

\subsection{The Auto--correlation Function}

We use the auto--correlation function (ACF) to calculate the time--scales of the fish. For scalar variables average over different individuals, such as $\Phi(t)$, $d_1(t)$, and $v_0(t)$, the ACF were calculated as

$$
\mathrm{ACF}(\tau) = \langle X(t) X(t + \tau) \rangle,
$$

\noindent where the bracket $\langle \cdots \rangle$ represents the average over different time points of the entire observation. The result functions were plotted in Fig.\ 2(a).

For the orientation ($\mathbf{o}_i = \mathbf{v}_i / |\mathbf{v}_i|$ where $\mathbf{v}$ is the velocity) of fish $i$, we calculate its ACF as

$$
\mathrm{ACF}(\tau)_i = \langle \mathbf{o}_i(t)
\cdot \mathbf{o}_i(t + \tau) \rangle_i,
$$

\noindent where the bracket $\langle \cdots \rangle$ represents the average over different time points along the trajectory of fish $i$, where the time points fall in one average window of 120 seconds. Then we calculate the average over different fish trajectores, to get one curve plotted in Fig.\ 2(b).

\subsection{The Radial Distribution Function}

We took extra care when calculating the radial radial distribution function ($g(r)$) due to the non-uniform distribution of the fish density, shown in Fig.\ 1 (c). Specifically, we calculate $g(r)$ as

$$
g(r) = \frac{P_\textrm{fish}(r)}{P_\textrm{id}(r)},
$$

\noindent 
instead of the counting the number of fish inside a spherical shell. Here $P_\textrm{fish}(r)$ is the probability of two fish having a pair-wise distance of $r$. The inhomogeneity is handled by biasing the spatial distribution of the ideal gas, \emph{i.e.}\ independent and random 3D points, to be identical to the spatial distribution of the fish. As a result, the ideal gas will have the same density distribution comparing with the fish, but lack the inherent pairwise structure of the fish. The term $P_\textrm{id}(r)$ is then the probability of a pair of particles at a separation $r$ for the spatially biased gas. Intuitively, the radio between the two reveals the pair-wise structure of the fish.

\section{Details of the Simulations}

For the inertial Vicsek model (IVM), we implemented the updating rule for the velocities in the main text,

\begin{widetext}
\begin{align*}
\mathbf{v}_{i}(t+1) = v_0 \Theta \left[(1 - \alpha)\; \underbrace{
v_{0} \mathcal{R}_{\eta}
\left[\Theta\left(
    \sum_{j \in S_{i}} \mathbf{v}_{j}(t)
\right)\right]
}_\text{Vicsek Model}
+ \alpha \mathbf{v}_i(t)
\right].
\label{eqIVM}
\end{align*}
\end{widetext}

\noindent The neighbours ($S_i$) of the $i$th particle are chosen to be all the particles whose distance is smaller than 1. After the velocities were updated from $\mathbf{v}(t)$ to $\mathbf{v}(t + 1)$, we update the positions to be $\mathbf{r}(t+1) = \mathbf{r}(t) + \mathbf{v}(t + 1)$.  The speed ($v_0$) of the particles was chosen to be 0.1. The number density of our simulation is set to 1 and we implemented a cubic periodic boundary condition. We run the simulation $10^5$ time steps before sampling, and we sampled another $10^5$ frames to get the results for analysis.

When calculating the orientational relaxation time {\mtau} for the simulated model, we selected the time when the orientational ACF reaches the value of $1/e$, since the initial ACF can not be easily fitted with an exponential decay. Changing this threshold value will shift the result, but the shift can always be compensated by varying the value of $\alpha$. All other {\descriptors} were calculated in the same way as the analysis of the experimental data.

\section{Numerical Verification of $lp \sim v_0 / \eta^{2}$}

\begin{figure*}
\centering
\includegraphics[width=120mm]{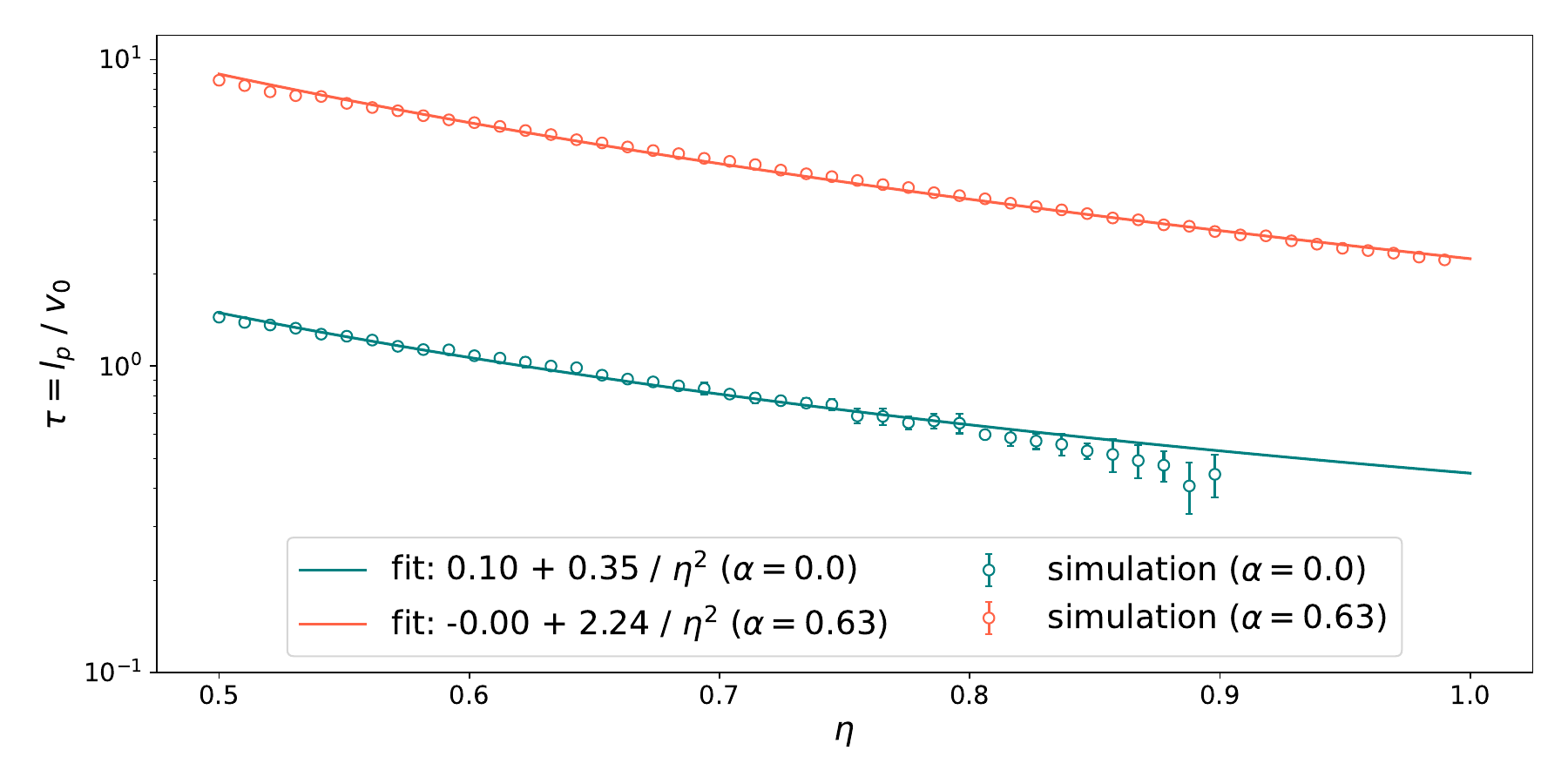}
\caption{\label{figLpScale}
The relationship between the orientational relaxation time ($\tau$) and the noise value ($\eta$) for the agents in the Vicsek model, in the absence of alignment interaction.
}
\end{figure*}

For the inertial Vicsek model, we verified the relationship betweel the persistence length ($l_p$) and noise ($\eta$), in the absence of alignment interactions. Figure~\ref{figLpScale} shows the relationship between the orientational relaxation time ($\tau = l_p / v_0$) of the Vicsek agents and the noise. The relaxation time is obtained by fitting the auto-correlation function of the orientation with an exponential decay. It is clear that the scaling relationship is good, for both the original Vicsek model ($\alpha=0$) and its inertial counterpart ($\alpha=0.63$).

\section{Comparing with Previous Results}

\begin{figure}
\centering
\includegraphics[width=\linewidth]{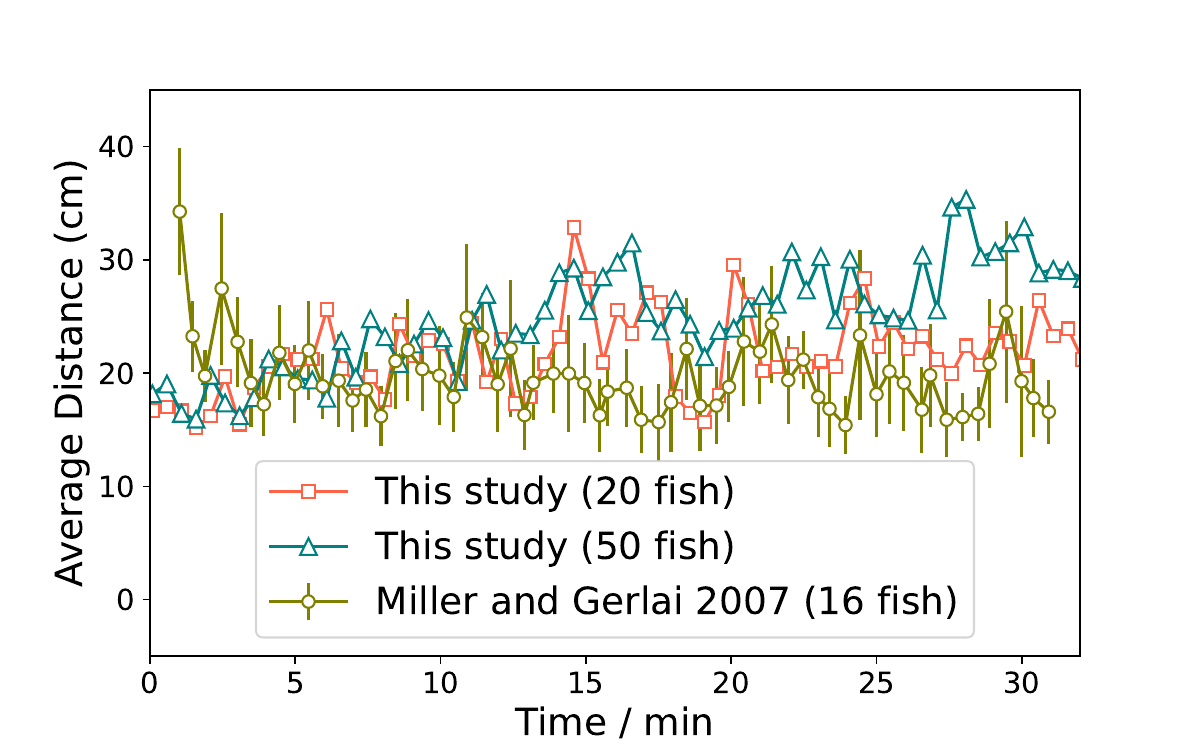}
\caption{The average distance values among the fish. Different symbols represent different experiments. For experiments in this study, the fish were young fish (body length $<$ 30 mm), and each scatter represent the average value over 1 second (15 frames). The error bar for Miller and Gerlai's data were from 8 different fish groups.}
\label{figCompareDist}
\end{figure}

We compared the averaged pairwise distance between the fish with the same results obtained by \citeauthor{miller2007} \cite{miller2007}. The comparison were presented in Fig.~\ref{figCompareDist}, and all the results were close to 20 cm, regardless of the group size.

\begin{figure}
\centering
\includegraphics[width=\linewidth]{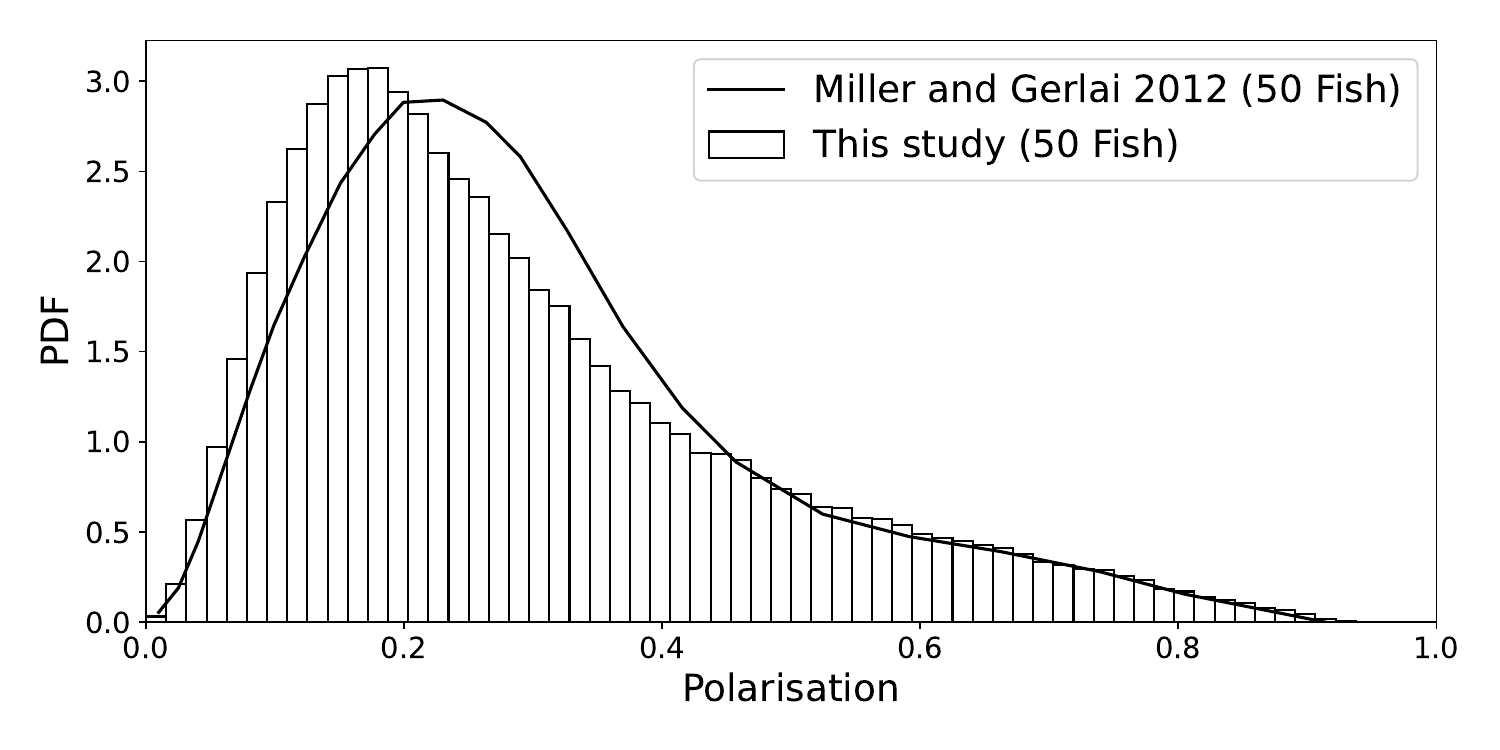}
\caption{\label{figComparePol}
The distribution of polarisation value of 50 zebrafish. The solid line was obtained from reference \cite{miller2012}. This data was originally reported as the probability mass function, and we adapted it to the probability density function to compare with our data.
}
\end{figure}

We also compared the distribution of the polarisation value of our fish (all groups) with the value reported by \citeauthor{miller2012} \cite{miller2012}. The result was plotted in Fig.~\ref{figComparePol}, presenting the distribution of polarisation values for 50 fish. The distribution from our data contains a peak in the disordered region ($\Phi \sim 0.2$) and a tail in the ordered region ($\Phi > 0.5$). Such distribution is very close to the result from \citeauthor{miller2012}.

\begin{acknowledgments}
We thank James G. Puckett, Christos C. Ioannou, and James Herbert for stimulating conversations. \yy{This work was carried out using the computational facilities of the Advanced Computing Research Centre, University of Bristol - http://www.bris.ac.uk/acrc/.}
\end{acknowledgments}


\begin{thebibliography}{10}

\bibitem{liu2019prl}
Liu G, Patch A, Bahar F, Yllanes D, Welch RD, Marchetti MC, et~al.
\newblock Self-{{Driven Phase Transitions Drive}}
  {{{\emph{Myxococcus}}}}{\emph{ Xanthus}} {{Fruiting Body Formation}}.
\newblock Physical Review Letters. 2019;122(24):248102.

\bibitem{cavagna2010}
Cavagna A, Cimarelli A, Giardina I, Parisi G, Santagati R, Stefanini F, et~al.
\newblock Scale-Free Correlations in Starling Flocks.
\newblock Proceedings of the National Academy of Sciences.
  2010;107(26):11865--11870.

\bibitem{makris2006}
Makris NC, Ratilal P, Symonds DT, Jagannathan S, Lee S, Nero RW.
\newblock Fish Population and Behavior Revealed by Instantaneous Continental
  Shelf-Scale Imaging.
\newblock Science. 2006;311(5761):660--663.

\bibitem{deneubourg1989}
Deneubourg JL, Goss S.
\newblock Collective Patterns and Decision-Making.
\newblock Ethology Ecology \& Evolution. 1989;1(4):295--311.

\bibitem{pitcher1982}
Pitcher TJ, Magurran AE, Winfield IJ.
\newblock Fish in Larger Shoals Find Food Faster.
\newblock Behavioral Ecology and Sociobiology. 1982;10(2):149--151.

\bibitem{trenchard2016}
Trenchard H, Perc M.
\newblock Energy Saving Mechanisms, Collective Behavior and the Variation Range
  Hypothesis in Biological Systems: {{A}} Review.
\newblock Biosystems. 2016;147:40--66.

\bibitem{hemelrijk2015}
Hemelrijk C, Reid D, Hildenbrandt H, Padding J.
\newblock The Increased Efficiency of Fish Swimming in a School.
\newblock Fish and Fisheries. 2015;16(3):511--521.

\bibitem{cavagna2017}
Cavagna A, Conti D, Creato C, Del~Castello L, Giardina I, Grigera TS, et~al.
\newblock Dynamic Scaling in Natural Swarms.
\newblock Nature Physics. 2017;13(9):914--918.

\bibitem{mendez-valderrama2018}
Méndez-Valderrama JF, Kinkhabwala YA, Silver J, Cohen I, Arias TA.
\newblock Density-Functional Fluctuation Theory of Crowds.
\newblock Nature Communications. 2018;9(1):3538.

\bibitem{grunbaum2005}
Grünbaum D, Viscido S, Parrish JK.
\newblock Extracting {{Interactive Control Algorithms}} from {{Group Dynamics}}
  of {{Schooling FIsh}}.
\newblock In: Cooperative {{Control}}; 2005. p. 103--117.

\bibitem{dorigo2006}
Dorigo M, Birattari M, Stutzle T.
\newblock Ant Colony Optimization.
\newblock IEEE computational intelligence magazine. 2006;1(4):28--39.

\bibitem{jia2015}
Jia Y, Wang L.
\newblock Leader–{{Follower Flocking}} of {{Multiple Robotic Fish}}.
\newblock IEEE/ASME Transactions on Mechatronics. 2015;20(3):1372--1383.

\bibitem{attanasi2014pcb}
Attanasi A, Cavagna A, Del~Castello L, Giardina I, Melillo S, Parisi L, et~al.
\newblock Collective Behaviour without Collective Order in Wild Swarms of
  Midges.
\newblock PLoS Computational Biology. 2014;10(7):e1003697.

\bibitem{ni2016}
Ni R, Ouellette NT.
\newblock On the Tensile Strength of Insect Swarms.
\newblock Physical Biology. 2016;13(4):045002.

\bibitem{gautrais2012}
Gautrais J, Ginelli F, Fournier R, Blanco S, Soria M, Chaté H, et~al.
\newblock Deciphering {{Interactions}} in {{Moving Animal Groups}}.
\newblock PLoS Computational Biology. 2012;8(9):e1002678.

\bibitem{bialek2012}
Bialek W, Cavagna A, Giardina I, Mora T, Silvestri E, Viale M, et~al.
\newblock Statistical Mechanics for Natural Flocks of Birds.
\newblock Proceedings of the National Academy of Sciences.
  2012;109(13):4786--4791.

\bibitem{sinhuber2017}
Sinhuber M, Ouellette NT.
\newblock Phase {{Coexistence}} in {{Insect Swarms}}.
\newblock Physical Review Letters. 2017;119(17):178003.

\bibitem{attanasi2014PRL}
Attanasi A, Cavagna A, Del~Castello L, Giardina I, Melillo S, Parisi L, et~al.
\newblock Finite-{{Size Scaling}} as a {{Way}} to {{Probe
  Near}}-{{Criticality}} in {{Natural Swarms}}.
\newblock Physical Review Letters. 2014;113(23):238102.

\bibitem{bricard2013}
Bricard A, Caussin JB, Desreumaux N, Dauchot O, Bartolo D.
\newblock Emergence of Macroscopic Directed Motion in Populations of Motile
  Colloids.
\newblock Nature. 2013;503(7474):95--98.

\bibitem{yan2016}
Yan J, Han M, Zhang J, Xu C, Luijten E, Granick S.
\newblock Reconfiguring Active Particles by Electrostatic Imbalance.
\newblock Nature Materials. 2016;15(10):1095--1099.

\bibitem{vicsek1995}
Vicsek T, Czirók A, Ben-Jacob E, Cohen I, Shochet O.
\newblock Novel {{Type}} of {{Phase Transition}} in a {{System}} of
  {{Self}}-{{Driven Particles}}.
\newblock Physical Review Letters. 1995;75(6):1226--1229.

\bibitem{couzin2002}
Couzin ID, Krause J, James R, Ruxton GD, Franks NR.
\newblock Collective {{Memory}} and {{Spatial Sorting}} in {{Animal Groups}}.
\newblock Journal of Theoretical Biology. 2002;218(1):1--11.

\bibitem{vandervaart2019}
van~der Vaart K, Sinhuber M, Reynolds AM, Ouellette NT.
\newblock Mechanical Spectroscopy of Insect Swarms.
\newblock Science Advances. 2019;5(7):eaaw9305.

\bibitem{giannini2020}
Giannini JA, Puckett JG.
\newblock Testing a Thermodynamic Approach to Collective Animal Behavior in
  Laboratory Fish Schools.
\newblock Physical Review E. 2020;101(6):062605.

\bibitem{puckett2018}
Puckett JG, Pokhrel AR, Giannini JA.
\newblock Collective Gradient Sensing in Fish Schools.
\newblock Scientific Reports. 2018;8(1):7587.

\bibitem{jolles2017}
Jolles JW, Boogert NJ, Sridhar VH, Couzin ID, Manica A.
\newblock {Consistent Individual Differences Drive Collective Behavior and
  Group Functioning of Schooling Fish}.
\newblock Current Biology. 2017 Sep;27(18):2862--2868.e7.

\bibitem{kim2017}
Kim OH, Cho HJ, Han E, Hong TI, Ariyasiri K, Choi JH, et~al.
\newblock Zebrafish Knockout of {{Down}} Syndrome Gene, {{DYRK1A}}, Shows
  Social Impairments Relevant to Autism.
\newblock Molecular Autism. 2017;8(1):50.

\bibitem{tang2020}
Tang W, Davidson JD, Zhang G, Conen KE, Fang J, Serluca F, et~al.
\newblock Genetic {{Control}} of {{Collective Behavior}} in {{Zebrafish}}.
\newblock iScience. 2020;23(3):100942.

\bibitem{pedersen2020}
Pedersen M, Haurum JB, Bengtson SH, Moeslund TB.
\newblock {3D-ZeF - A 3D Zebrafish Tracking Benchmark Dataset.}
\newblock CVPR. 2020;.

\bibitem{butail2011}
Butail S, Paley DA.
\newblock {Three-dimensional reconstruction of the fast-start swimming
  kinematics of densely schooling fish}.
\newblock Journal of The Royal Society Interface. 2011 Nov;9(66):77--88.

\bibitem{wang2017}
Wang SH, Liu X, Zhao J, Liu Y, Chen YQ.
\newblock {3D tracking swimming fish school using a master view tracking first
  strategy}.
\newblock In: Proceedings - 2016 IEEE International Conference on
  Bioinformatics and Biomedicine, BIBM 2016. Shanghai University of Engineering
  Science, Shanghai, China. IEEE; 2017. p. 516--519.

\bibitem{cachat2011}
Cachat J, Stewart A, Utterback E, Hart P, Gaikwad S, Wong K, et~al.
\newblock Three-{{Dimensional Neurophenotyping}} of {{Adult Zebrafish
  Behavior}}.
\newblock PLoS ONE. 2011;6(3):e17597.

\bibitem{saberioon2016}
Saberioon MM, Cisar P.
\newblock {Automated multiple fish tracking in three-Dimension using a
  Structured Light Sensor}.
\newblock Computers and Electronics in Agriculture. 2016 Feb;121(C):215--221.

\bibitem{shelton2020}
Shelton DS, Shelton SG, Daniel DK, Raja M, Bhat A, Tanguay RL, et~al.
\newblock Collective Behavior in Wild Zebrafish.
\newblock Zebrafish. 2020;17(4):243--252.

\bibitem{kelley2013}
Kelley DH, Ouellette NT.
\newblock Emergent Dynamics of Laboratory Insect Swarms.
\newblock Scientific Reports. 2013;3(1):1073.

\bibitem{stowers2017}
Stowers JR, Hofbauer M, Bastien R, Griessner J, Higgins P, Farooqui S, et~al.
\newblock {Virtual reality for freely moving animals}.
\newblock Nat Meth. 2017 Aug;14(10):995--1002.

\bibitem{ling2019}
Ling H, Mclvor GE, van~der Vaart K, Vaughan RT, Thornton A, Ouellette NT.
\newblock Costs and Benefits of Social Relationships in the Collective Motion
  of Bird Flocks.
\newblock Nature Ecology \& Evolution. 2019;3(6):943--948.

\bibitem{yang2022}
Yang Y, Turci F, Kague E, Hammond CL, Russo J, Royall CP.
\newblock Dominating Lengthscales of Zebrafish Collective
  Behaviour;18:e1009394.

\bibitem{mwaffo2015}
Mwaffo V, Anderson RP, Butail S, Porfiri M.
\newblock A Jump Persistent Turning Walker to Model Zebrafish Locomotion.
\newblock Journal of The Royal Society Interface. 2015;12(102):20140884.

\bibitem{hansen2013}
Hansen JP, McDonald IR.
\newblock Theory of Simple Liquids: with Applications to Soft Matter.
\newblock Elsevier Science; 2013.

\bibitem{ginelli2016}
Ginelli F.
\newblock The {{Physics}} of the {{Vicsek}} Model.
\newblock The European Physical Journal Special Topics.
  2016;225(11-12):2099--2117.

\bibitem{ballerini2008}
Ballerini M, Cabibbo N, Candelier R, Cavagna A, Cisbani E, Giardina I, et~al.
\newblock {Empirical investigation of starling flocks: a benchmark study in
  collective animal behaviour}.
\newblock Animal Behaviour. 2008 Jul;76(1):201--215.

\bibitem{ling2019nc}
Ling H, Mclvor GE, Westley J, Vaart K, Vaughan RT, Thornton A, et~al.
\newblock {Behavioural plasticity and the transition to order in jackdaw
  flocks}.
\newblock Nat Comms. 2019 Nov;p. 1--7.

\bibitem{silverberg2013}
Silverberg JL, Bierbaum M, Sethna JP, Cohen I.
\newblock Collective {{Motion}} of {{Humans}} in {{Mosh}} and {{Circle Pits}}
  at {{Heavy Metal Concerts}}.
\newblock Physical Review Letters. 2013;110(22):228701.

\bibitem{vandervaart2020}
van~der Vaart K, Sinhuber M, Reynolds AM, Ouellette NT.
\newblock Environmental Perturbations Induce Correlations in Midge Swarms.
\newblock Journal of The Royal Society Interface. 2020;17(164):20200018.

\bibitem{delaney2002}
Delaney M, Follet C, Ryan N, Hanney N, Lusk-Yablick J, Gerlach G.
\newblock Social {{Interaction}} and {{Distribution}} of {{Female Zebrafish}} (
  {{{\emph{Danio}}}}{\emph{ Rerio}} ) in a {{Large Aquarium}}.
\newblock The Biological Bulletin. 2002;203(2):240--241.

\bibitem{miller2012}
Miller N, Gerlai R.
\newblock From Schooling to Shoaling: Patterns of Collective Motion in
  Zebrafish ({{Danio}} Rerio).
\newblock PLoS ONE. 2012;7(11):e48865.

\bibitem{perez-escudero2017}
Pérez-Escudero A, de~Polavieja GG.
\newblock Adversity magnifies the importance of social information in
  decision-making.
\newblock Journal of The Royal Society Interface. 2017;14(136):20170748.

\bibitem{miller2007}
Miller N, Gerlai R.
\newblock Quantification of Shoaling Behaviour in Zebrafish ({{Danio}} Rerio).
\newblock Behavioural Brain Research. 2007;184(2):157--166.

\bibitem{miller2013}
Miller N, Greene K, Dydinski A, Gerlai R.
\newblock Effects of Nicotine and Alcohol on Zebrafish ({{Danio}} Rerio)
  Shoaling.
\newblock Behav Brain Res;240:192--196.

\bibitem{heras2019}
Heras FJH, Romero-Ferrero F, Hinz RC, de~Polavieja GG.
\newblock Deep Attention Networks Reveal the Rules of Collective Motion in
  Zebrafish.
\newblock PLOS Computational Biology. 2019;15(9):e1007354.

\bibitem{tunstrom2013}
Tunstrøm K, Katz Y, Ioannou CC, Huepe C, Lutz MJ, Couzin ID.
\newblock Collective {{States}}, {{Multistability}} and {{Transitional
  Behavior}} in {{Schooling Fish}}.
\newblock PLoS Computational Biology. 2013;9(2):e1002915.

\bibitem{mora2011}
Mora T, Bialek W.
\newblock {Are Biological Systems Poised at Criticality?}
\newblock J Stat Phys. 2011 06;144(2):268 -- 302.

\bibitem{lawrence2018}
Lawrence EA, Kague E, Aggleton JA, Harniman RL, Roddy KA, Hammond CL.
\newblock {The mechanical impact of col11a2 loss on joints; col11a2 mutant
  zebrafish show changes to joint development and function, which leads to
  early-onset osteoarthritis.}
\newblock Philos Trans R Soc Lond B Biol Sci. 2018 Sep;373(1759).

\bibitem{westerfield2000}
Westerfield M.
\newblock {The Zebrafish Book: A Guide for the Laboratory Use of Zebrafish
  (Danio Rerio)}.
\newblock University of Oregon Press; 2000.

\bibitem{hartley2003}
Hartley R, Zisserman A.
\newblock Multiple View Geometry in Computer Vision.
\newblock Cambridge Books Online. {Cambridge University Press}; 2003.

\bibitem{cavagna2008AB}
Cavagna A, Giardina I, Orlandi A, Parisi G, Procaccini A, Viale M, et~al.
\newblock The {{STARFLAG}} Handbook on Collective Animal Behaviour: 1.
  {{Empirical}} Methods.
\newblock Animal Behaviour. 2008;76(1):217--236.

\bibitem{ouellette2006}
Ouellette NT, Xu H, Bodenschatz E.
\newblock A Quantitative Study of Three-Dimensional {{Lagrangian}} Particle
  Tracking Algorithms.
\newblock Experiments in Fluids. 2006;40(2):301--313.

\bibitem{xu2008}
Xu H.
\newblock Tracking {{Lagrangian}} Trajectories in Position–Velocity Space.
\newblock Measurement Science and Technology. 2008;19(7):075105.

\bibitem{mwaffo2017}
Mwaffo V, Butail S, Porfiri M.
\newblock In-Silico Experiments of Zebrafish Behaviour: Modeling Swimming in
  Three Dimensions.
\newblock Scientific Reports. 2017;7(1):39877.

\bibitem{rosa2020}
Rosa LV, Costa FV, Canzian J, Borba JV, Quadros VA, Rosemberg DB.
\newblock Three- and Bi-Dimensional Analyses of the Shoaling Behavior in
  Zebrafish: {{Influence}} of Modulators of Anxiety-like Responses.
\newblock Progress in Neuro-Psychopharmacology and Biological Psychiatry.
  2020;102:109957.

\bibitem{parrish1997}
Parrish JK, editor.
\newblock Animal Groups in Three Dimensions.
\newblock {Cambridge Univ. Press}; 1997.

\bibitem{yushi2021}
Yang Y. yangyushi/FishPy: bird catcher.
\newblock Zenodo; 2021.
\newblock Available from: \url{https://doi.org/10.5281/zenodo.4735711}.

\end{thebibliography}
\end{document}